# Beyond SDLC: Process Modeling and Documentation Using Thinging Machines

Sabah Al-Fedaghi
*salfedaghi@yahoo.com*
Computer Engineering Department, Kuwait University, Kuwait

**Summary**
The software development life cycle (SDLC) is a procedure used to develop a software system that meets both the customer's needs and real-world requirements. The first phase of the SDLC involves creating a conceptual model that represents the involved domain in reality. In requirements engineering, building such a model is considered a bridge to the design and construction phases. However, this type of model can also serve as a basic model for identifying business processes and how these processes are interconnected to achieve the final result. This paper focuses on process modeling in organizations, per se, beyond its application in the SDLC when an organization needs further documentation to meet its growth needs and address regular changes over time. The resultant process documentation is created alongside the daily operations of the business process. The model provides process visualization and documentation to assist in defining work patterns, avoiding redundancy, or even designing new processes. In this paper, a proposed diagrammatic representation models each process using one diagram comprising five actions and two types of relationships to build three levels of depiction. These levels consist of a static description, events, and the behavior of the modeled process. The viability of a thinging machine is demonstrated by re-modeling some examples from the literature.

## 1. Introduction

The software development life cycle (SDLC) is a procedure used to produce a software system that meets both the customer's needs and real-world requirements. The first phase of the SDLC involves creating a conceptual model that represents the involved domain in reality. In requirements engineering, building such a model is considered a bridge to the design and construction phases. However, outside of the SDLC context, a similar type of model can serve as a basic apparatus for identifying activity flows (task steps) in an organization's processes (e.g., information technology [IT], business, physical, or cyber procedures) and how these processes are interconnected to achieve the final result. In this context, processes refer to various workflows such as transforming materials, delivering services, or handling data/information revealing how the organization works from the inside [1]. According to the International Organization for Standardization (ISO) 9001, an organization must organize and maintain information regarding various processes relevant to its systems. For example, the ISO 9001/2015 auditing procedure requires one to show organized "system documents with the most updated information and have it available and within reach for management and employees who need to refer to it" [2].

This paper concentrates on process documentation in organizations based on process *modeling*. We propose a non-flowchart-based modeling methodology with which to construct process documentation for processes in an existing system. The diagrammatic model represents each process using one diagram comprising five actions and two types of relationships to build three levels of representation: a static description, events, and the behavior of the modeled process. The viability of a thinging machine (TM) is demonstrated by re-modeling some examples from the literature.

1.1 What Is Process Documentation?

According to economist W. E. Deming, "If you can't describe what you are doing as a process, you don't know what you're doing." An organization's *processes* define how products and services are developed, manufactured, and delivered [3]. In this context, processes address an organization's internal aspects and structures, and they cover managerial and technical aspects, particularly the deployed IT infrastructure. A process may be specified in text or diagram form, or it may be written in a formal language. Typically, flowchart-based descriptions are used to develop a model and the so-called workflow specifications. In addition to flowcharts, many other techniques and tools are used in conjunction with processes, such as DFDs, UML, and Petri nets.

Process *modeling* provides the visualization and documentation of processes to assist in defining work patterns, avoiding redundancy, or even designing new processes. Process documentation is a form of documentation schematics that describe how to perform a business process. Its functions include facilitating communication, a guide for understanding enterprise processes and a tool for refining productivity and aligning with new goals. [4].

Process *documentation* is produced alongside the daily operations of a business process, rather than within the requirements engineering phase of the SDLC. It is what is called a conceptual model during the SDLC—a diagrammatic representation "created in an agreed modeling language" [3]. In physical systems, process documentation is developed when a new process is set up.

1.2 Problem and Solution

An organization may lack process documentation for various reasons, including a deficiency in requirements development or the adaptation of a finished software system product. Documenting and updating a business process are difficult tasks. In many situations, documentation likely will not be done unless it is mandatory (e.g., legally). Many enterprises use documents as high-level memos. According to Chaffee [5], "None of them [documents] could be given to a new hire to use in training or to help them understand how to do their job. They don't actually help employees do their jobs better." Documenting processes may include many processes, each with many pages, resulting in a large volume of seldom used papers. One way to meet this challenge is standardization (e.g., ITIL). Another approach, as proposed in this paper, is to adopt a new type of diagrammatic-based modeling that reduces the central parts of documentation to a single diagram for each process.

1.3 Examples of the New Approach to Process Modeling

Because our method focuses on modeling a process using a single three-level diagram, it is beneficial to clarify the types of diagrams proposed in this paper right from the start. Hence, Figs. 1 and 2 provide samples of process models.

Fig. 1 models a process performed by a network engineer working in an actual environment. The process includes a client computer; multiple switches, routers, servers, security elements, users, and protocols; and many other subprocesses. The network administrator (circle 1 in the figure) logs into his/her active directory and performs tasks such as (a) adding a new user name and assigning privileges and (b) designating the client computer name by logging into the user's computer (2). Similar processes are followed for deleting a user name and a computer name. Accordingly, an account and its password are given to the user who logs in (3). Complete details of this process documentation can be found in [6].

The above example involves an IT process. Process documentation may involve a physical process, such as the process of filling cargo oil, which involves tanks that bring crude oil from oil-field centers to vessels in ports, as shown in Fig. 2. Of course, the modeling may include the IT processes as part of the physical process. More details about this project can be found in [7].

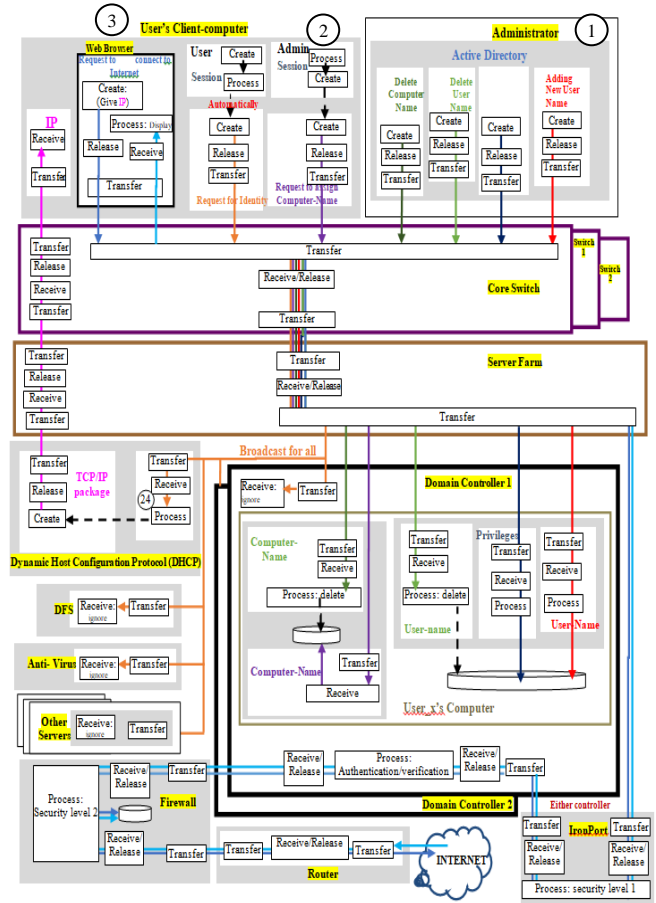

Fig. 1. The process of adding or removing a user from the network.

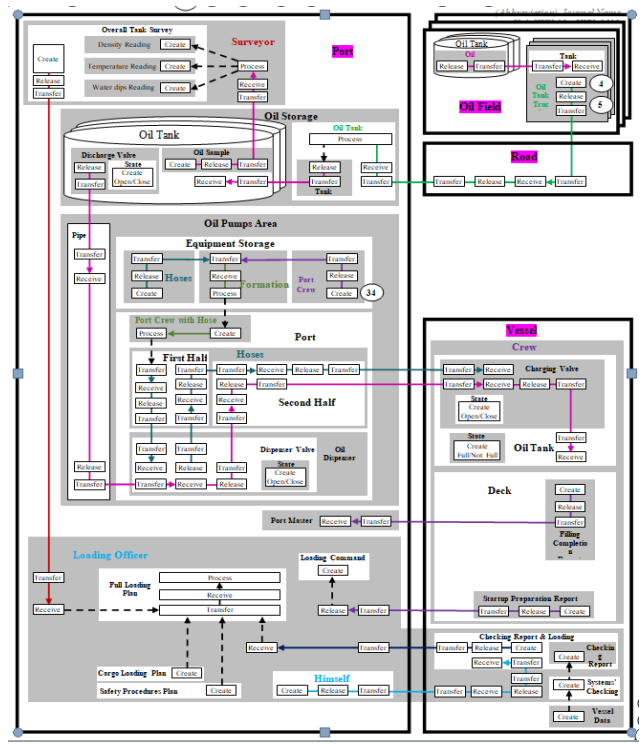

Fig. 2. The process of cargo oil filling.

## 2. Documenting Processes in Organizations

Process documentation is described as a roadmap that shows how work is done and includes a graphical depiction of the organization's processes [8]. According to René et al. [1], an example of process documentation is a building permit procedure that involves several services of the public administration at various levels. Administrations need a way to understand one another's methods of handling administrative procedures through a common language (e.g., Business Process Model and Notation [BPMN] 2.0). Process documentation is difficult [8]. A 2017 survey of 1,500 global IT leaders reported that "poor process documentation is one of the five biggest IT failures of all time" [9]. For example, when it comes to network and information security threats, organizational processes—"the way in which things are done"—create serious cybersecurity vulnerabilities to threats due to internal process failures, including poor process flow and poor process documentation [10].

Flowchart-based methods are usually adopted to describe processes (e.g., classical flowcharts, UML activity diagrams, flow diagrams, and DFDs). These notations lack genericity, which refers to a limited set of elementary actions. An alternative tool exists. For example, Figs. 1 and 2 in the introduction were developed using what is called the TM.

### 2.1. TM Modeling

To illustrate how the TM differs from other techniques, we introduce Six Sigma *process maps* [11], which are shown in Fig. 3. According to the Six Sigma Institute [11], "Process maps help characterize the functional relationships between various inputs and outputs… Process mapping is a graphics technique for dissecting a process by capturing and integrating the combined knowledge of all persons associated with the process." The Six Sigma Institute [11] provides Fig. 3 as an example of such a graphics technique, which involves the *process of using a telephone*. Fig. 4 shows the TM representation of this process, which is described in more detail in this paper.

In Fig. 4, first, the person (circle 1) receives the telephone, in the sense that he/she possesses a telephone (2), and processes (uses) it (3) by dialing its numbers. This causes a ringing sound (4), which is the result of sending signals to the other side. The reaction is either receiving a response (5) or silence (6). If a response is received, then this initiates a reply (7), thus starting a conversation (8). Note that Fig. 4 utilizes only two generic verbs: "process" and "create." In general, the TM utilizes five such actions (create, process, release, transfer, and receive), which are discussed later in this paper. The solid arrows in Fig. 4 denote flows, and the dashed arrows indicate triggering. Fig. 5 shows the division of the TM model for using a telephone into seven events. Fig. 6 shows the behavior of the modeled process. In contrasting the TM representation with the Six Sigma flowchart, the TM shows far richer semantics and simple specification.

After this outline of the types of proposed diagrammatic notations, we return to the issues related to process documentation in general.

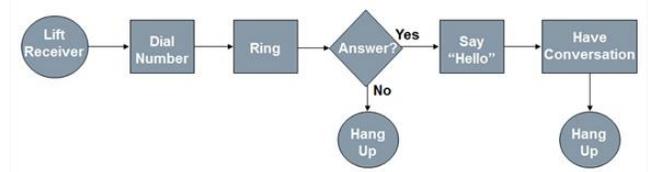

Fig. 3. Six Sigma process flowchart (adopted from [11]).

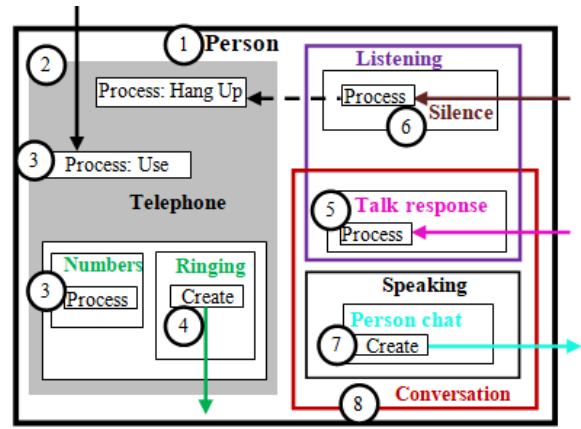

Fig. 4. The process description of using a telephone.

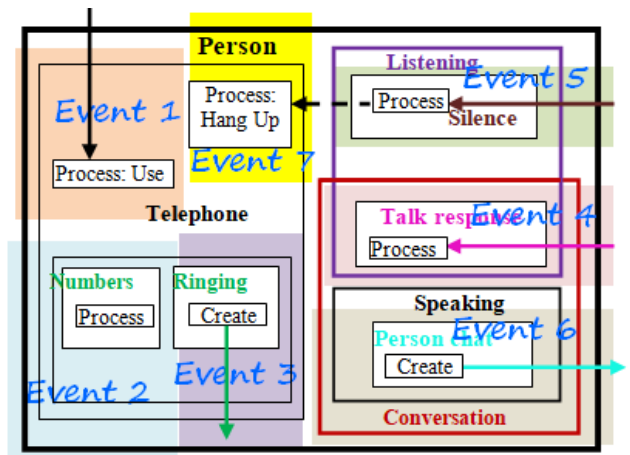

Fig. 5. Events in the process description of using a telephone.

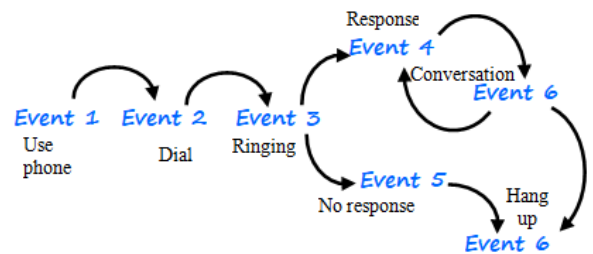

Fig. 6. Behavioral model of the process of using a telephone.

2.2. Advantages of Process Documentation

- Process documentation provides a foundation on which to develop information and communication technologies, maintain information systems, and enable IT units and organizational units to work together effectively [12].
- An organization's ability to reorganize resources dynamically and adjust to changing environmental settings is essential for business success [4]. It is essential to monitor change within an organization. Consequently, process documentation is a tool for understanding the different phases of the business process life cycle.
- Process documentation offers employees insights into their job roles, the entire business process, and their interdependencies with others. This facilitates communication because it can only be done jointly among several employees [12].
- A documented process may serve many purposes. It can provide a baseline for analysis and the opportunity to improve organizational cohesion, identify bottlenecks and inefficiencies, provide training to employees, and provide a means for objective evaluation [8].

2.3. Relationship between Process Documentation and Modeling

As mentioned previously, modeling can assist during the requirements phase of software development and, in fact, throughout the entire SDLC process. This paper focuses on using process modeling in organizations, per se, beyond its application during the SDLC, when an organization needs further documentation to meet its growth needs and address regular changes over time. Time and growth are usually accompanied by complexity. Documentation, re-documentation, and extra documentation are steps for handling such an increase in complexity. Process modeling describes the actors, activities, event sequences, and manipulated objects. Different organizational departments can utilize a process model. For example, in the BPMN context, a model can be constructed for each task separately, whether it is a so-called human task, service task, or user task.

Chow et al. [13] propose three distinct values delivered through the use of process modeling: further (a) efficiency (by reducing operational costs, improving productivity, etc.), (b) control (ensuring compliance, improving visibility, and managing the process outcomes), and (c) agility (adapting quickly to changing world conditions, e.g., having the speed to create and change processes).

## 3. How to Document Processes

"Thinging" is a term borrowed from the German philosopher Heidegger [14], who suggested that *thinging* expresses how a thing *things*—for example, how an organization gathers or ties together its constituent parts. The term "machine" comes from viewing any organization simultaneously as a (holistic) thing and as a (operational) machine. The TM reduces millions of actions in English to just five actions: create, process, release, transfer, and receive (see [15] for references). For example, a customer *submitting* an order can be expressed as follows:

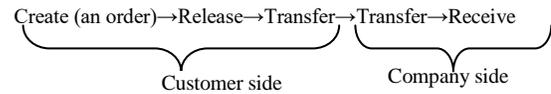

The customer *creates* an order and may put it on hold for a while, so it is in the *release* state (e.g., creating an email but not sending it); then, the customer transfers the order. *Transfer* on the company side denotes the company's input (e.g., a port or incoming/outgoing tray) before the item or message is actually *received* (e.g., an order may arrive but never be received due to it being misplaced). Continuing with the stream of actions above, the company *processes* the order; (a) *triggers* the creation of an invoice, which is sent to the customer; and then (b) forwards the order to the inventory. Such an internal company process is modeled as follows:

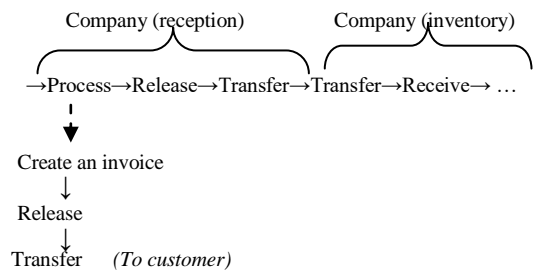

Note that the two models of the processes given in the introduction—the models of adding to or removing from a network and cargo oil filling—are constructed based on the repeated usage of these five generic verbs. TM notations are simple because they include just the (a) five verbs, (b) a solid arrow denoting the flow, and (c) a dashed arrow denoting triggering. Additionally, the TM can model the dynamic features (the execution of the modeled process) and time events (as demonstrated in the telephoning process discussed previously) without using any additional notation.

In this paper, we propose using the TM as a documentation tool for existing IT, cyber-physical, or physical processes within organizations. We propose developing one diagram (including static dynamic and behavior models) for each process. The rest of this paper applies the TM to larger case studies to exhibit different features of TM modeling.

## 4. Case Study 1 of Process Documentation

According to Rábová [16], the UML is OMG's most frequently used specification and is how the world models not only an application's structure, behavior, and architecture but also business processes, workflows, and data structures. An activity diagram is a "good way to show how different workflows are managed, how they start, go and stop and show many different decision paths that can be taken from start to finish" [16]. Rábová [16] provides a sample case of the "order making" process, modeled as the activity diagram shown in Fig. 7.

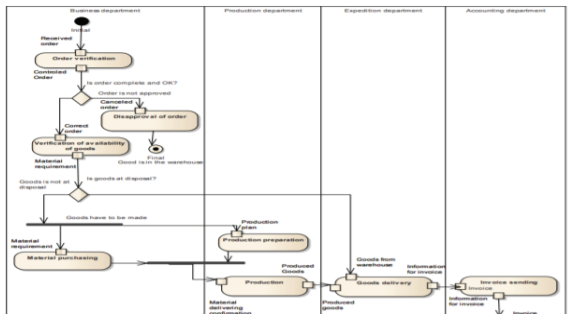

Fig. 7. Partial view of the UML activity diagram of order making (partially from [14]).

In the example, activities relating to particular entities within the model are placed within swimlanes to indicate their association. In activity diagrams, we can map, measure, and interpret all aspects of workflows and business documents.

The purpose of Fig. 7 is not to discuss the activity model but to present a general view of the flowchart-based modeling used. The details of this "order making" process can be found in [16]. We will translate this process into TM representation.

Fig. 8 shows the corresponding TM model, developed to the best of our understanding, of the model involved in [16].

Fig. 8 contains five TM machines: the customer (circle 1), management (2; of the system), the inventory (3), the supplier (4; of raw materials), and production (5; of products). We assume that the order structure, which is analogous to the UML conceptual class, contains three attributes of the order: the product number, date, and customer information (6). Each attribute includes the action *create* to indicate that the attribute must be filled with a value. Filling these values results in the order being *created* (7). The order flows (8) to the management system, where it is processed (9).

- If the order is not OK (10; e.g., if it is incomplete), then this triggers (11) the creation (12) of a cancel notification, which flows to the customer (13).
- If the order is OK (14), then this causes the order to be sent to the inventory (15 and 16).

In the inventory, the order is processed (17) to trigger the creation of a reply (18) to management indicating whether the product is available (19) or unavailable (20).

- If the product is available in the inventory, then this triggers (21) the creation (22) of an invoice, which flows to the customer (23).

Accordingly, the customer reacts by creating a payment (24; the upper-left corner of the figure). The payment is sent to the management system (25), where it is processed (26).

- If the payment is not OK (26), it is not clear in the given activity diagram how this should be handled; hence, we leave it because no further action can be taken here.

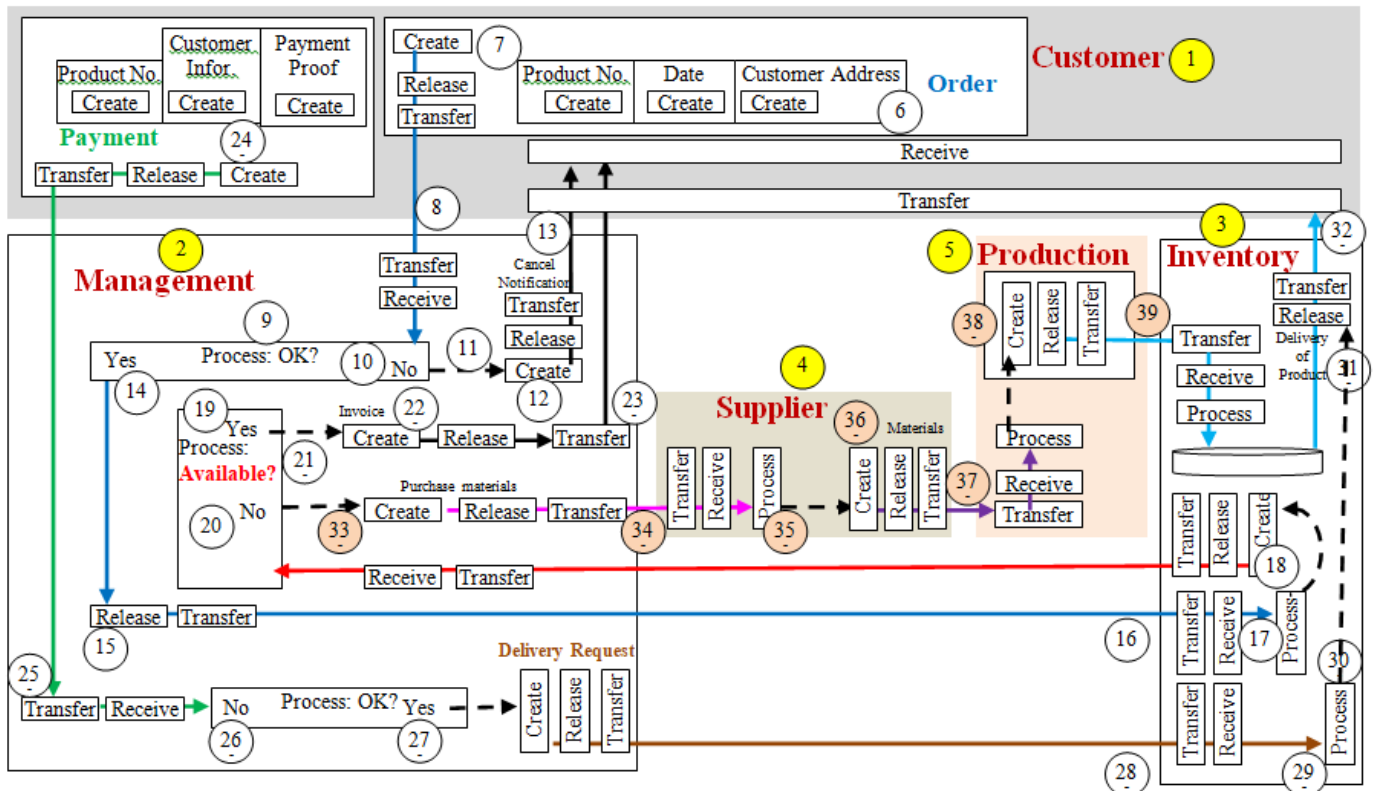

Fig. 8. Static TM model of the order-making process.

- If the payment is OK, then a delivery request is created (27) and sent to the inventory (28).

In the inventory, the instruction is processed (29; the bottom-right corner) to trigger (30) the product's release (31) to the customer (32).

Returning to the case in which the product is not available in the inventory (20), this triggers the creation (33; the red circle in the middle of the figure) of a purchase-material order, which flows to the supplier (34), where it is processed (35). Accordingly, the supplier creates the materials (36), which flow to production (37), where they are processed to create the product (38), which flows to the inventory (39).

The behavior of the order-making process is defined by identifying the event regions in the TM model to produce the dynamic description. An event in the TM is defined in terms of a time and region (sub-diagram of Fig. 8). For example, Fig. 9 shows the event *The customer sends a payment to management*.

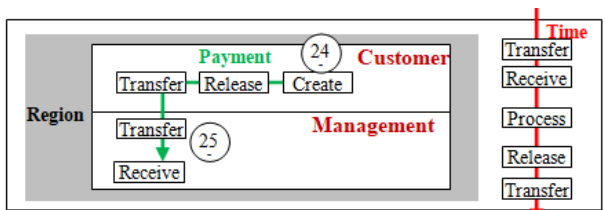

Fig. 9. The event *The customer sends a payment to management*.

For simplicity's sake, we use only the region to denote the event. Accordingly, Fig. 10 shows the following events, where event i is denoted by $E_i$.

$E_1$: A customer creates an order and sends it to the management system.
$E_2$: Management processes the order.
$E_3$: The order is not OK; hence, a cancel notification is sent to the customer.
$E_4$: The order is OK; hence, it is sent to the inventory.
$E_5$: The inventory reports on the availability/nonavailability of the ordered product to management.
$E_6$: Management processes the availability/nonavailability of the ordered product.
$E_7$: The ordered product is available; hence, an invoice is sent to the customer.
$E_8$: The customer sends a payment.
$E_9$: Management processes the payment.
$E_{10}$: The payment is not OK (further actions not specified in [14]).
$E_{11}$: The payment is OK; hence, a delivery request is sent to the inventory.
$E_{12}$: The inventory sends the product to the customer.
$E_{13}$: The ordered product is unavailable in the inventory.
$E_{14}$: Materials are purchased from the supplier.
$E_{15}$: The supplier sends the materials to production.
$E_{16}$: Production makes the product and supplies it to the inventory.

Fig. 11 shows the behavioral model in terms of the chronology of events.

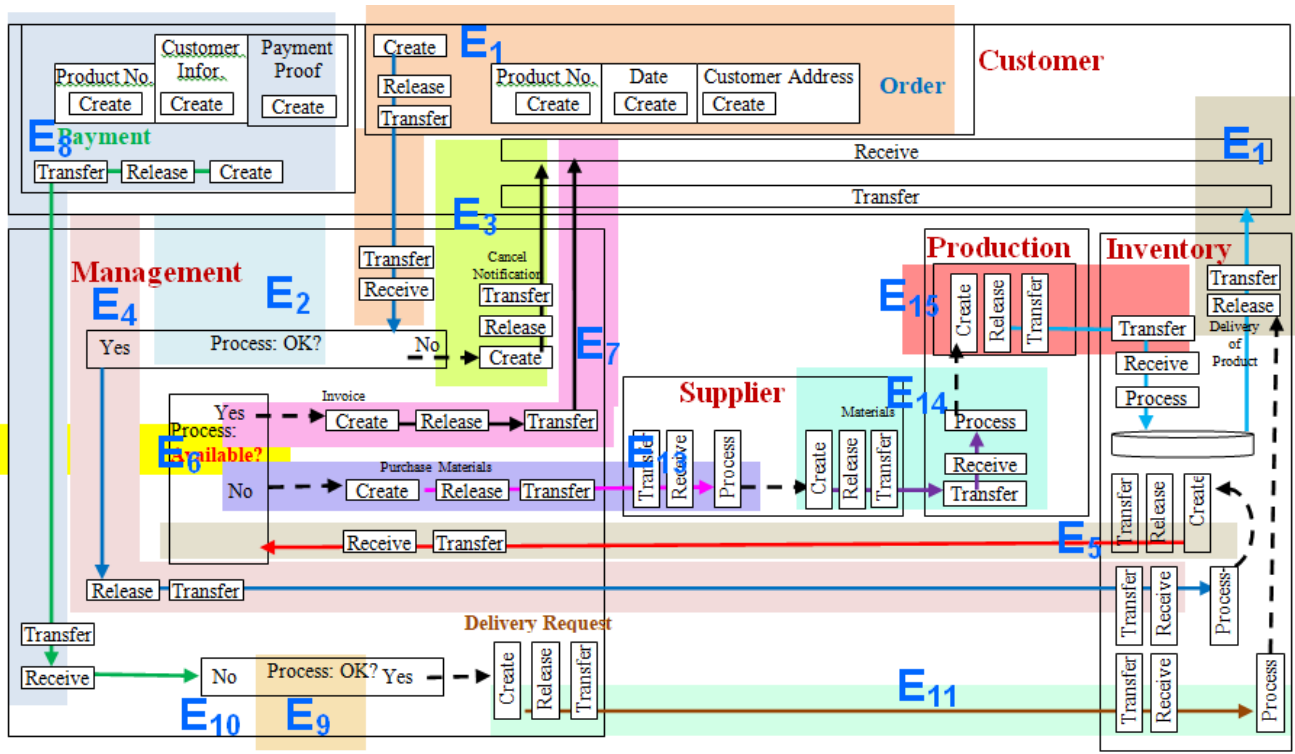

Fig. 10. Events TM model of the order-making process.

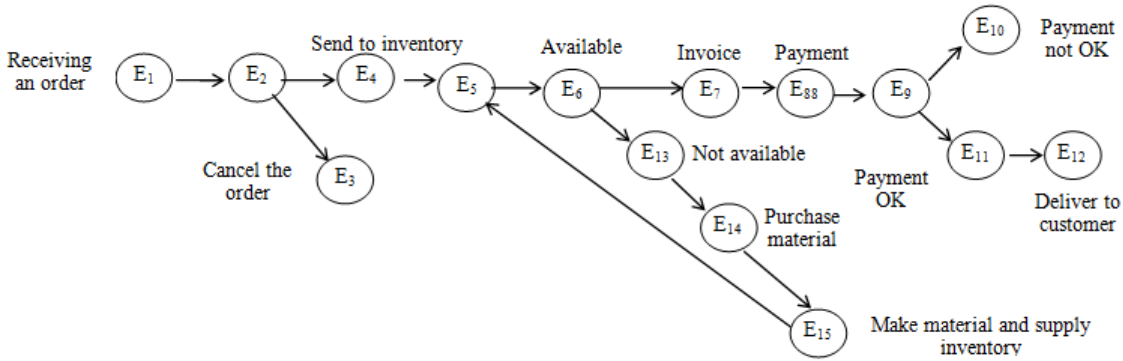

Fig. 11. The TM behavioral model of the order-making process.

## 5. Case Study 2

Schenker [8] models a *product-order* process (https://www.site.uottawa.ca/~bochmann/ELG7187C/CourseNotes/BehaviorModeling/Petri-nets/index.html) similar to the order-making process presented in the previous section, using a UML activity diagram, a use case, and Petri nets, as shown in Figs. 12, 13, and 14. This case study provides an opportunity to contrast the TM diagrams with these modeling forms. We observe the following:

- The activity diagrams, use cases, and Petri net models lack the notion of the generic action of the TM model. Thus, in general, millions of "activities" may be proportional to verbs in English.
- Activity diagrams, use cases, and Petri net models fail to distinguish sharply between a static description and event and behavioral specification. Thus, actions and time-oriented events are condensed into the same type of flat diagram. In the TM, events are identified at a second level of the diagrammatic representation to facilitate defining the chronology of events.

The Petri net model is distinctive as a formal notation. Hence, a TM mathematical foundation must be developed, either independently or based on Petri nets. Some efforts have already been made in this direction.

## 6. Documenting Physical Multiparty Processes

According to Meroni and Plebani [17], many business processes now cross the boundaries of single organizations, thus becoming multiparty processes. This also affects the goods involved in the process and means that multiple organizations may manipulate and alter the process when it is executed. Meroni and Plebani [17] and others (e.g., [18]) explore the monitoring of multiparty business processes, which involves monitoring smart physical objects (they are aware of their own conditions). Meroni and Plebani [17] present a case study concerning the process of the shipment of dangerous goods. The actors are a manufacturer, a customer, and a truck driver that are involved when potentially explosive chemicals must be delivered from the manufacturer to the customer. The delivery process is organized according to the BPMN [19] model shown in Fig. 15.

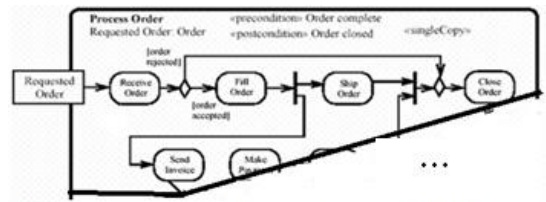

Fig. 12. The product-order problem in an activity diagram (from [8]).

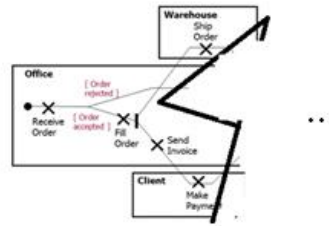

Fig. 13. The product-order problem in a use case (from [8]).

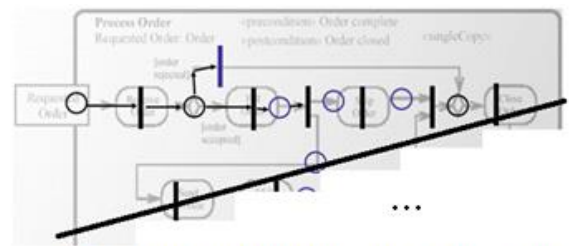

Fig. 14. The product-order process in Petri nets (from [8]).

This case study provides an opportunity to apply the TM to document a physical process, as shown in Fig. 16, in which the following occur:
- The manufacturer (circle 1) has storage (2), which includes smart tanks (3) containing sensors (4).
- To start the filling operation, a tank is taken from the storage (5) to attach (6) with a hose (7).

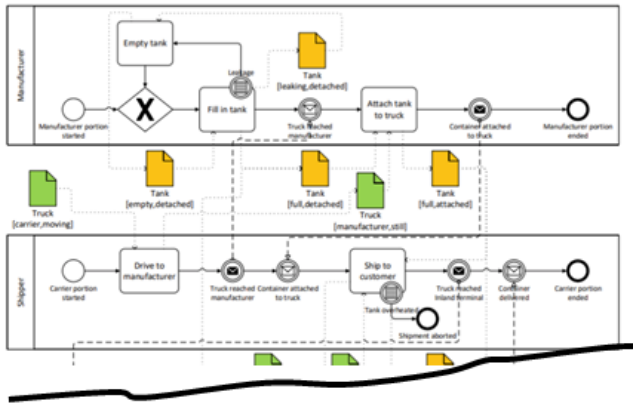

Fig. 15. Partial view of the BPMN diagram (from [17]).

We assume that the hose is the source of the dangerous goods that will fill the tank.

- Accordingly, a "fill unit" is formed from the tank and hose (8), and the filling process begins.

- During the filling process, two possibilities are as follows:
  - The sensor in the tank (10) issues a leakage alarm (11), which the manufacturer's observers receive (12), triggering (13) the detachment of the tank from the hose (14).

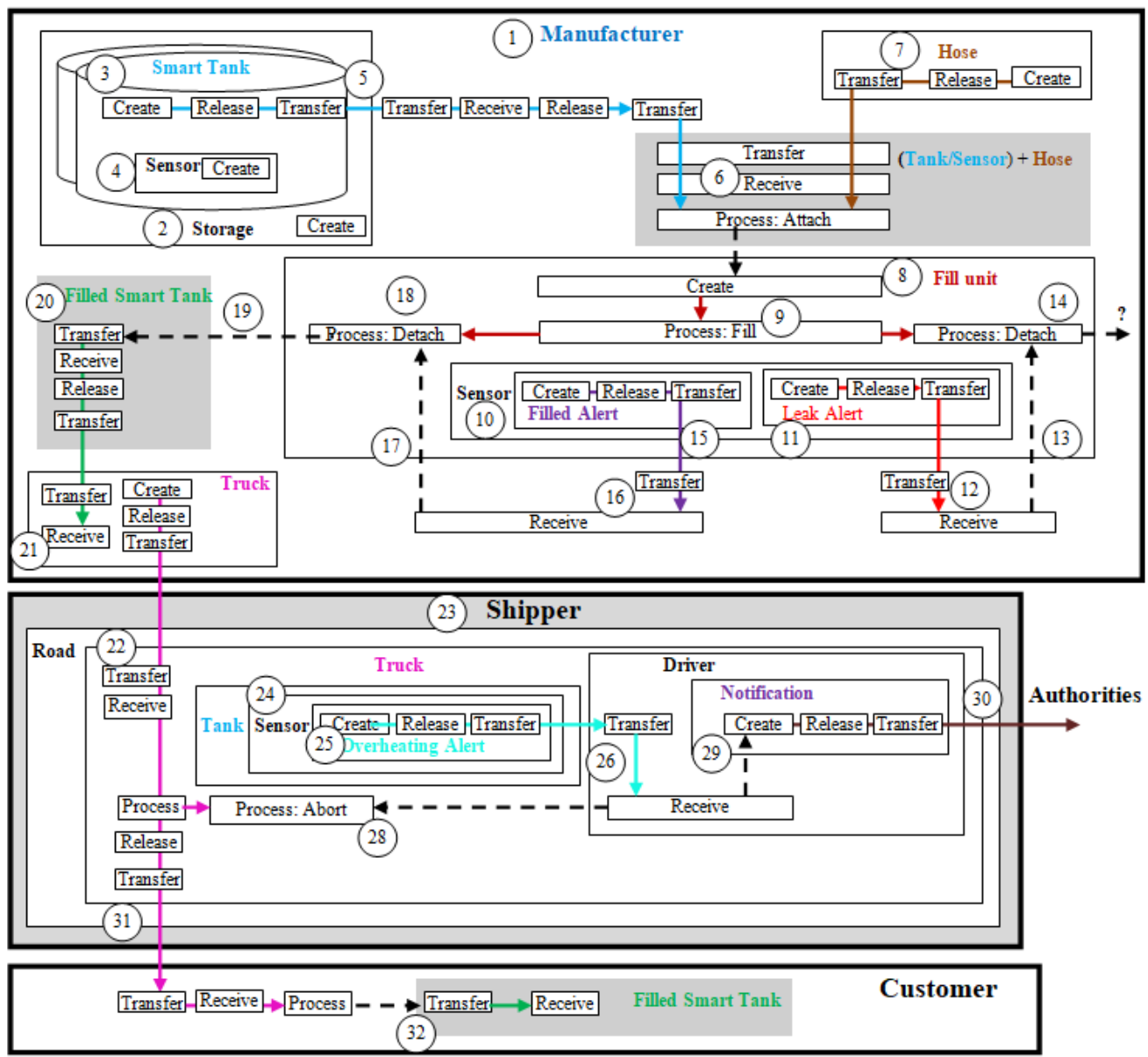

Fig. 16. The TM static model for monitoring smart physical objects.

- The sensor in the tank issues a completion alarm (15), which the observers receive (16), triggering (17) the detachment of the tank from the hose (18).
- The completion of the filling results in the appearance (19) of a filled tank (20), which moves to the truck parked in the manufacturer's area (21).
- The truck carrying the tank begins its journey on the road (22) and thus becomes the responsibility of the driver (23).
- The sensor in the tank (24) may issue an overheating alarm (25) in the tank. This alarm is received (26) by the driver (27). In this case, the driver aborts the mission (28) and contacts the authorities (29 and 30).
- If no alarm occurs on the road, the truck reaches the customer (31), where the tank is delivered (32).

Fig. 17 shows the TM event model with the following events, where $E_i$ denotes event i.
$E_1$: A smart tank is brought from the storage.
$E_2$: The tank is sent to the place where it is attached to the filling hose.
$E_3$: The hose is brought to be attached to the tank.
$E_4$: The tank and the hose are attached.
$E_5$: A filling unit is created.
$E_6$: The filling process is performed.
$E_7$: A leak alert occurs, which the control unit observes.
$E_8$: The tank is detached, and a procedure (not described in the original problem) is performed.

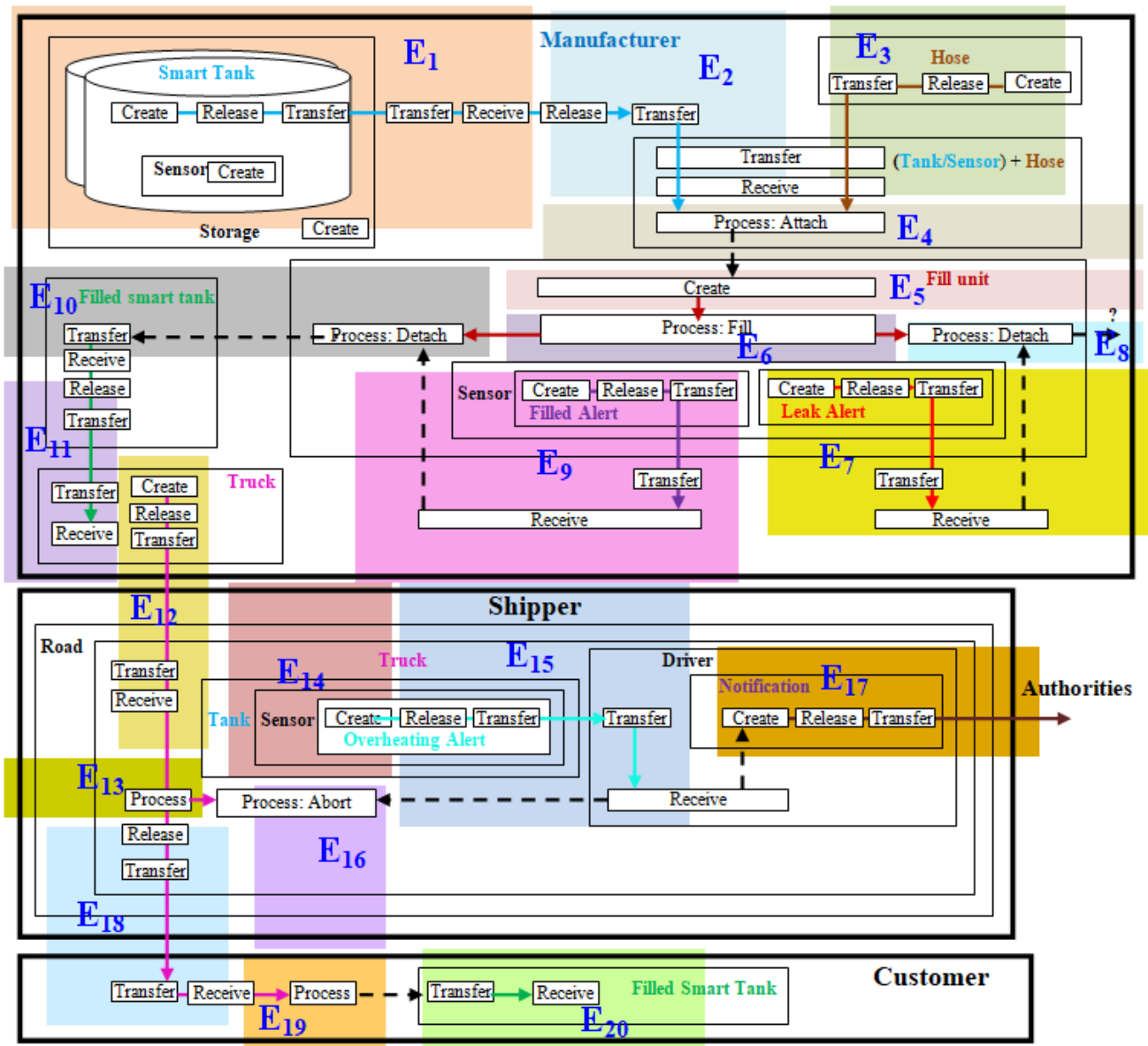

Fig. 17. The TM events model of monitoring smart physical objects.

$E_9$: A filled-tank alarm occurs, and the control unit observes it.
$E_{10}$: The filled tank is detached.
$E_{11}$: The filled tank is loaded on the truck.
$E_{12}$: The truck moves to the road.
$E_{13}$: The truck is driven on the road, on its way to the customer.
$E_{14}$: A leak alert occurs.
$E_{15}$: The driver hears the alert.
$E_{16}$: The driver aborts the mission.
$E_{17}$: The driver notifies the authorities.
$E_{18}$: The truck reaches the customer.
$E_{19}$: The tank is unloaded from the truck.

Fig. 18 shows the behavioral model in terms of the chronology of events for the physical objects being moved.

## 7. Conclusion

The focus of this paper has been on documenting and modeling business processes, whether IT, business, physical, or cyber processes. Often, these models either are not constructed or are produced but not systematically maintained, thus creating so-called "pollution" in the organization's process-model repositories [20]. We have proposed a non-flowchart-based modeling methodology (TM) with which to construct process documentation for a currently existing system.

### 7.1. Advantages of the TM Model

A diagrammatic TM model represents each process using one diagram with five actions and two types of arrows to build three levels of representation: a static description, events, and the behavior of the modeled process (Fig. 19). Such an explicit distinction between a description, events, and behavior is unique to the TM, in comparison with other flat, one-level tools, such as flowcharts, UML diagrams, and DFDs. We claim that the TM would be a useful tool for managing processes in the same way that network engineers utilize network maps. The diagram's size is compensated for by the simplicity of the repeated usage of its five actions and two types of arrows. The viability of the TM is demonstrated by re-modeling some examples from the literature.

### 7.2. Alleged Complexity of the TM Model

According to a referee, the TM diagrams are not very helpful. If the objective is to provide a better notation for modeling such that it provides a basis for rigorous analysis, reasoning, and documentation, then these diagrams are not suitable. Each TM diagram is very busy.

All such diagrams include a group of rectangles, circles, and arrows, and they seem to be arbitrarily arranged on a page. The purpose of modeling is to make complex phenomena easier to understand and more intuitive. The behavior models do not help understand much either. They are a group of circles with arbitrary arrows attached to them. What is an arrow? A transition? A data flow? A control flow? An edge connecting one node to another?

Many previous papers (e.g., [21-22] and their references) have explained the basic notions of the TM model. They include the following:
- **Things** (roughly correspond to UML objects). A thing is a more general notion than an object. It is based on the German philosopher Heidegger's [14] notion of a sharp distinction between objects and things and his claim that the word "thing" is richer and more meaningful.
- **Generic actions**: create, process, release, transfer, and receive. TM claims that diagrammatic modeling can be accomplished by using only these actions.
- **Flows** (denoted by arrows). The TM flows denote the conceptual movement of things among generic actions in their context.

These are all notations of the TM model. Events are specified over sub-diagrams of the static diagram, so this does not require additional notation. The term *transition* is not used in TM. It comes from the state diagram terminology, which does not distinguish between the static and events levels. The state diagram is a one-level representation that does not involve time, a notion necessary for events.

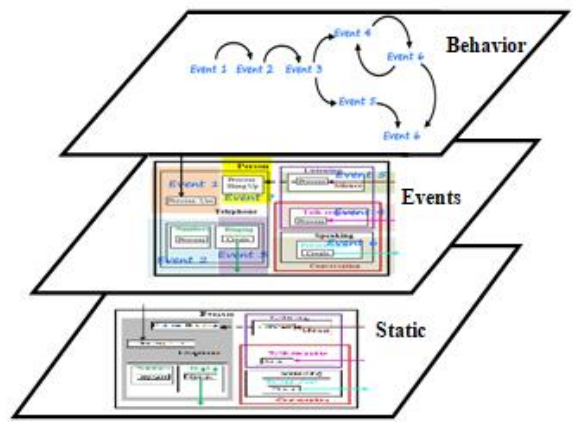

Fig. 19. The three-level TM model

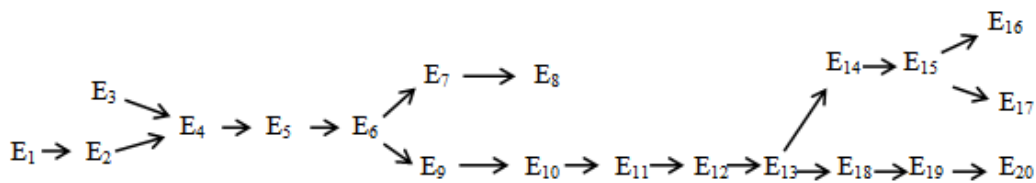

Fig. 18. The behavioral model.

Accordingly, the claim that the TM diagram is a group of rectangles, circles (circles are used for mere explanation of different spots on the diagram, but they are not in TM notation), and arrows, which seem to be arbitrarily arranged on a page, is not justified. The five generic actions are *systematically* repeated based on a TM that specifies the permitted flow. They enrich and reduce the specification to a generic level (no more basic actions).

Additionally, it is claimed that the TM is a complex tool to be used for modeling, say, in comparison with the standard modeling language UML. The apparent complexity is the result of completeness of specification. Nevertheless, the TM diagram can be reduced greatly by several simplification processes.

Fig. 20 shows a first-level simplification of the static TM model of the order-making process (Fig. 8). The simplification is accomplished by assuming that the direction of arrows in the TM diagram eliminates the need for the generic actions release, transfer, and receive. This is analogous to achieving simplicity of computer hardware by eliminating the (a) input/output port (transfer), the input buffer (receive), and the output buffer (release) of a computer (e.g., hardwired connection to the computer storage and CPU). Fig. 21 presents further simplification by eliminating the create action under the assumption, for example, of triggering leading to the creation of a new flow. Fig. 22 shows yet another simplification by eliminating the "word" process and dashed arrows.

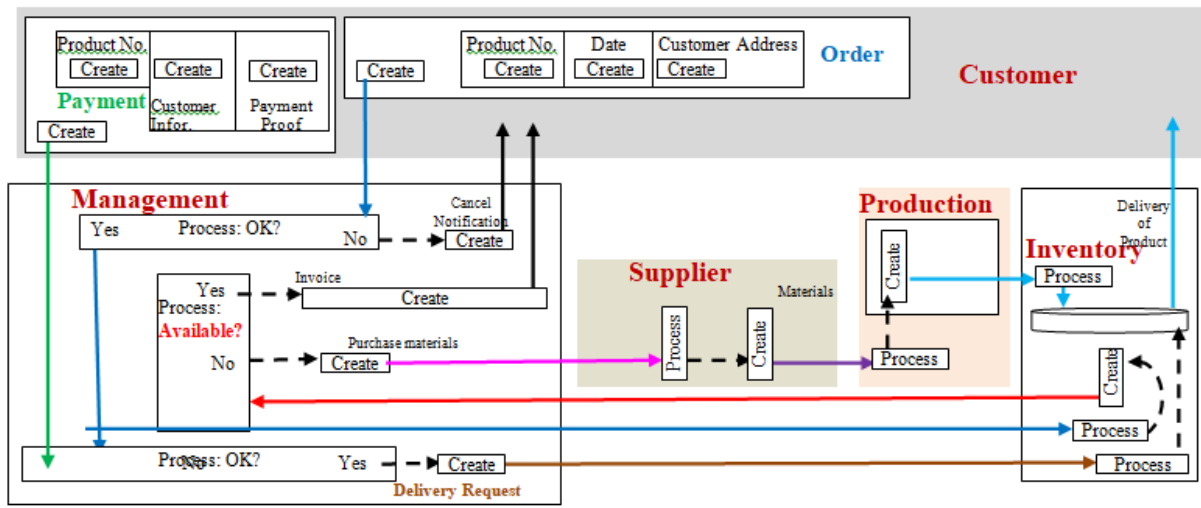

Fig. 20 First-level simplification of the static TM model of the order-making process.

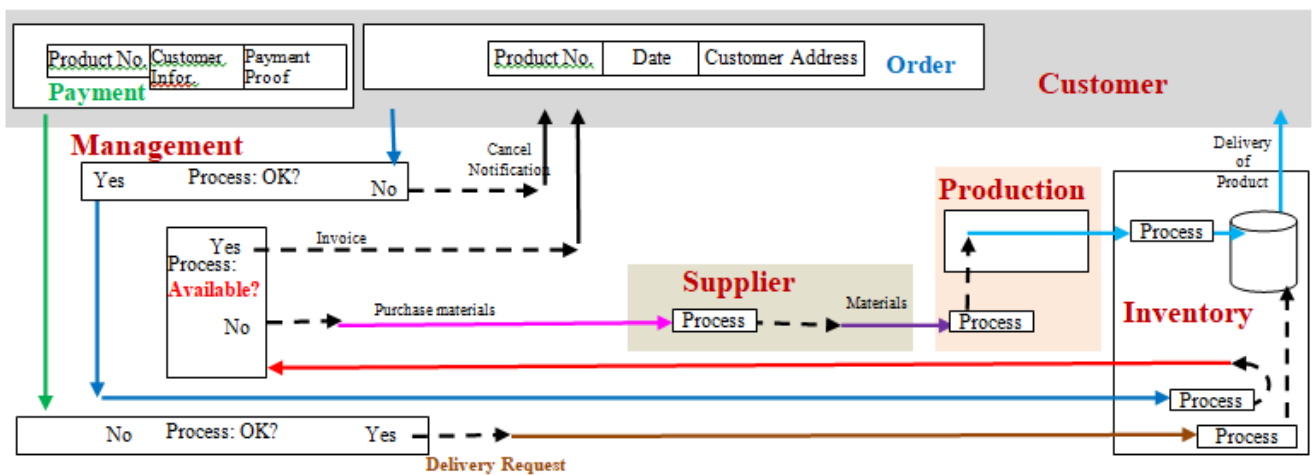

Fig. 21. Second-level simplification of the static TM model of the order-making process.

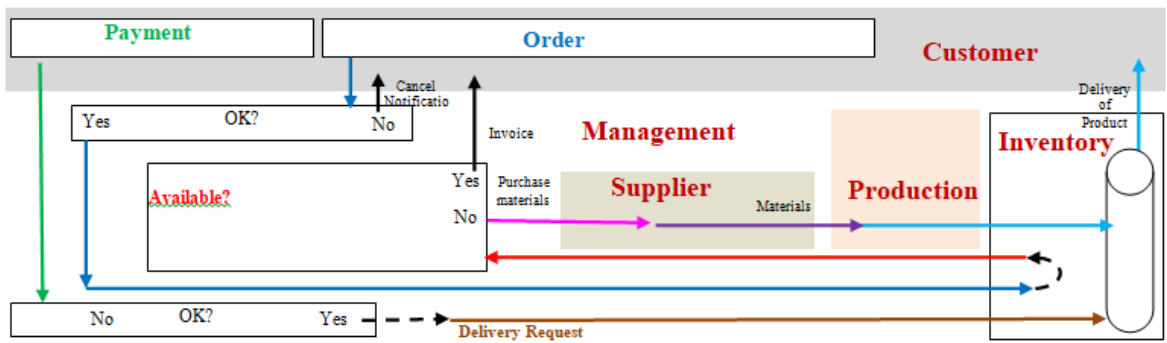

Fig. 22 Third-level simplification of the static TM model of the order-making process.

We can go further in this simplification until we reach a figure with mere boxes that denotes functionalities. Each of the simplified figures can be supplemented with an explanation of its process. It is important to note that these simplified diagrams are based on the *complete description of Figure 8,* which specifies all basic processes that can be coded into software.

Accordingly, a TM diagram is an engineering diagram that will be realized as a tangible product (software). Figs. 23 and 24 show two samples of engineering schemata from two different scientific fields. Both are obviously very complex descriptions (far more than any TM diagram). Does such a complexity imply inadequate models because it is not easier to understand the electronic/aeronautic phenomena? Why is most of the world built upon such descriptions of systems? How do electronic/aeronautic engineers specify these complex circuits? What are the initial conceptual steps that lead to such specification?

Complexity is a relative term. However, when two representations involve the same level of abstraction, we can say that one of them is more complex than the other. UML is known for its complexity because it involves 14 models, each with different notations. In spite of the wide adoption of the object-oriented approach and UML as the most common modeling paradigm, "The use of object-concepts in conceptual modeling has not been widely adapted. A main reason is that there are no generally accepted semantics of these concepts as conceptual modeling elements" [23]. In object-oriented modeling, "The basic concepts are tightly interrelated and cannot be easily taught and learned in isolation" [24]. This complexity is intrinsic to object orientation and cannot be removed [25].

In general, according to Sedrakyan et al. [26], "There is a certain degree of difficulty in understanding a system represented by means of UML diagrams." A survey of UML practitioners [27] shows that UML class diagrams are not fully used for further software development, either for code generation or for documentation.

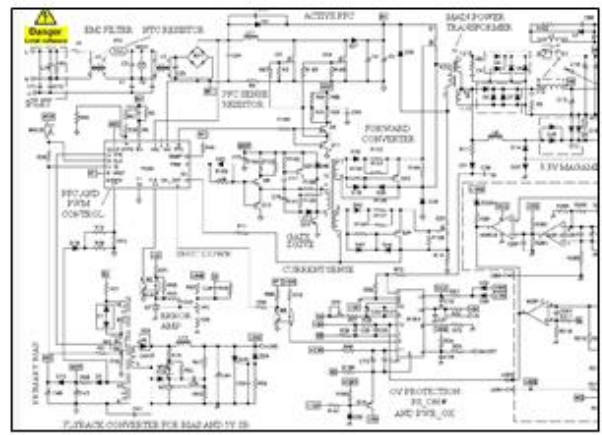

Fig. 23 Electronic system schemata (https://www.smpspowersupply.com/atx-power-supply.html)

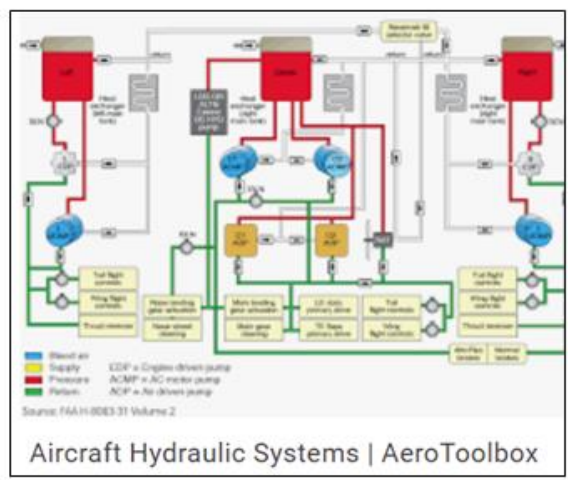

Fig. 24 Aircraft hydraulic system drawing (https://www.flight-mechanic.com/hydraulic-system-components-part-one/)

It has been reported that some commercial industries find modeling to be cumbersome and slow down productivity [27]. "For such projects, it makes sense to use UML as [merely] a sketch and have your model contain some architectural diagrams and a few class and sequence diagrams to illustrate key points" [28].

According to Aguirre-Urreta et al. [29], papers comparing diagrammatic conceptual models (e.g., entity relationship and UML/object-oriented modeling techniques) in the published literature, although vibrant, have often yielded equivocal findings. In this context, Houy et al. [30] state that model understandability remains ambiguous, and research results on model understandability are hardly comparable and partly imprecise. One way to contrast conceptual models is through experimentation.

For example, Valaski et al. [31] used eight professionals and 80 students to evaluate the expressiveness of UML and OntoUML. The point here is that it is very difficult to present a detailed comparison between UML and TM modeling, especially because the latter is still a mere proposed approach. Achieving a reasonable level of comparability at this stage of development involves modeling the same problem in UML and TM and contrasting the diagrammatic representations side by side in a way that can be grasped by all stakeholders.

On the other hand, TM's apparent complexity appears as the result of repeatedly using the five generic actions create process, release, transfer, and receive. It has, as mentioned previously, few notations in comparison with UML, which involves many notations of the 14 types of modeling. Thus, the notion of complexity cannot be used to dismiss TM modeling.

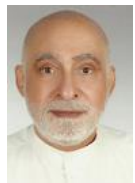

**Sabah S. Al-Fedaghi** is an associate professor in the Department of Computer Engineering at Kuwait University. He holds an MS and a PhD from the Department of Electrical Engineering and Computer Science, Northwestern University, Evanston, Illinois, and a BS from Arizona State University. He has published more than 350 journal articles and papers in conferences on software engineering, database systems, information ethics, privacy, and security. He headed the Electrical and Computer Engineering Department (1991–1994) and the Computer Engineering Department (2000–2007). He previously worked as a programmer at the Kuwait Oil Company. Dr. Al-Fedaghi has retired from the services of Kuwait University on June 2021.